\documentclass[aps,amsfonts,nofootinbib,twocolumn]{revtex4}
\usepackage{graphicx}
\usepackage{dcolumn}
\usepackage{bm}

\begin{document}


\title{Photonic Crystal
Nanobeam Cavity Strongly Coupled to the Feeding Waveguide}

\author{Qimin Quan}
\email{quan@fas.harvard.edu}
\author{Parag B Deotare}
\author{Marko Loncar}

\affiliation{
School of Engineering and Applied Sciences, Harvard University, Cambridge, Massachusetts 02138, USA
}%

\begin{abstract}
A \emph{deterministic} design of an ultrahigh Q, wavelength scale mode volume photonic crystal nanobeam
cavity is proposed and experimentally demonstrated. Using this approach, cavities with Q$>10^6$ and
on-resonance transmission T$>$90\% are designed. The devices fabricated in Si and capped with low-index polymer, have Q=80,000 and T=73\%.
This is, to the best of our knowledge, the highest transmission measured in deterministically designed,
wavelength scale high Q cavities.
\end{abstract}

\maketitle

\begin{figure}
\centering
\includegraphics[width=9cm,height=6cm]{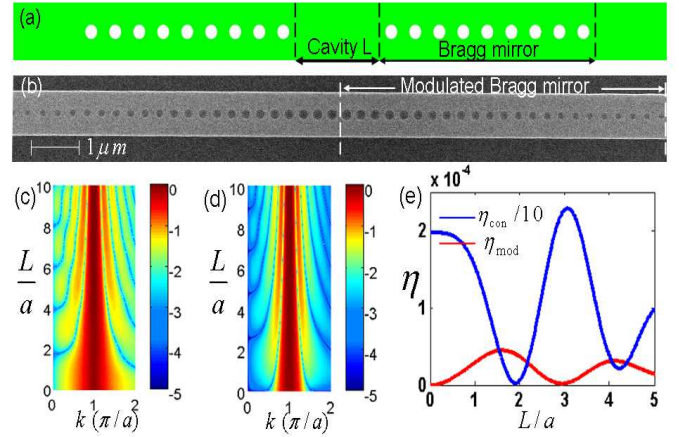}
\caption{\small (a)Schematic of nanobeam PhC with conventional Bragg
mirror. (b)SEM image of a silicon nanobeam PhC with "modulated
Bragg mirror". (c)\&(d) $\mathrm{log}_{10}|\mathrm{FT}(H)|$ for a model cavity with different cavity length
L(normalized to the period) and two different types of mirror that support: c)exponential attenuation with $\kappa=0.6$ and d)Gaussian
attenuation with $\sigma=0.1$. (e)The portion of the fourier
components that are inside the lightline assuming an index propotion
of 2.5 between the waveguide and the surroundings.}\label{model}
\end{figure}

Photonic crystal (PhC)\cite{Yablonovitch87}\cite{John87} cavities,
with quality(Q) factors over million and wavelength scale mode
volumes, are widely applied in the fields that range from quantum
information processing and nonlinear optics, to biomedical sensing.
The fourier space analysis\cite{Vuckovic02}\cite{Katik02}, the
multi-pole cancelation\cite{Johnson01}, and the mode matching
mechanism\cite{watts02}\cite{lalanne01}\cite{lalanne04} have been
developed to explain the origin of high Qs. The design of PhC
cavities, however, is typically based on extensive parameter search and
optimization\cite{Noda03}-\cite{Notomi08}, also known as intuitive
design. The large computational cost, in particular the computation
time, needed to perform the simulation of high-Q cavities make this
trial based method inefficient. Inverse engineering design, in which the physical structure is optimized
by constructing specific target
functions and constraints, was also
proposed\cite{geremia02}\cite{Burger04}. A design
recipe based on the desired
field distribution is proposed in \cite{Dirk05}. In this letter, we
propose and experimentally demonstrate a \emph{deterministic} method
to design an ultrahigh Q, sub-wavelength scale mode volume, PhC
nanobeam cavity(Figure.\ref{model}) that is strongly
coupled to the feeding waveguide(i.e. near unity on resonance
transmission). The design approach is deterministic in the sense
that it does not involve any trial-based hole shifting, re-sizing
and overall cavity re-scaling to ensure ultra-high Q cavity. Moreover, the final cavity resonance
has less than 2\% deviation from a predetermined frequency. Our design
method requires only computationally inexpensive, photonic band
calculations (e.g. using plane wave expansion method), and is simple to
implement.

The Q factor of a PhC nanobeam cavity can be maximized by reducing
the out-of plane scattering($Q_{\mathrm{sc}}$) due to the coupling
to the radiation modes. As shown
previously\cite{Vuckovic02}\cite{Dirk05}, scattered power
($P_{\mathrm{sc}}$) can be expressed as an integral of spatial
fourier frequencies within a light cone, calculated over the surface
above the cavity: $P_{\mathrm{sc}}\propto\int_{<\small
\mathrm{lightcone}}dk(|FT(H_z)|^2+|FT(E_z)|^2)$. The integral is
minimized when major fourier components are tightly localized (in
k-space) at the edge of the first Brillioun zone\cite{Katik02}. We
start by considering the ideal field distribution on this surface
which would minimize $P_{\mathrm{sc}}$. A general property of these
nanobeam cavities is that it consists of the waveguide region of
length L, that supports propagating modes, surrounded by infinitely
long Bragg mirror on each side(Figure.\ref{model}a). Without the loss of generality, we
consider the TE-like cavity mode with Hz as a major field component.
In the case of conventional periodic Bragg mirror, evanescent field
inside the mirror can be expressed as $\sin(\beta_{\mathrm{Bragg}}
x)\exp(-\kappa x)$, where $\kappa$ is attenuation constant. The
cavity field inside the waveguide region can be represented as
$\sin(\beta_{\mathrm{wg}} x)$. As mentioned above, scattering loss
decreases in mirror section when $\beta_{\mathrm{Bragg}}=\pi/a$,
while phase matching between mirror and waveguide\cite{lalanne01},
$\beta_{\mathrm{Bragg}}=\beta_{\mathrm{wg}}$, minimizes the
scattering loss at cavity-mirror interface. The spatial fourier
transform of such cavity field is approximately  a Lorentzian in the
vicinity of $\pi/a$. As proposed in \cite{Noda03}, spatial
frequencies within light-cone can be minimized (Q maximized) if the
field attenuation inside the mirror has a Gaussian shape
$~\sin(\beta_{\mathrm{Bragg}} x)\exp(-\sigma x^2)$. In Figure
\ref{model}(c)\&(d), the fourier space spectrums of the cavity modes
with exponential and with Gaussian attenuation are shown respectively. The
fraction of the energy associated with spatial harmonics within the
light-cone, $\eta$, for both cases is shown in
Figure.\ref{model}(e). It can be seen that cavity with the Gaussian
attenuation has more than an order of magnitude smaller $\eta$.

The
preferred Gaussian attenuation can be obtained by making $\kappa$ a
linearly increasing function of the position within the mirror($\kappa=\sigma x$). We
name such mirror "modulated Bragg mirror". At the same time, to
prevent scattering, the oscillating part of the field,
$\sin(\beta_{\mathrm{mod}}x)$, should have constant
$\beta_{\mathrm{mod}}$ throughout the mirror. This condition is
satisfied only if each segment of modulated Bragg mirror has the
same length ("periodicity") $a$, and if the operating frequency is
kept inside the bandgap of each segment of the modulated Bragg
mirror. In the case of dielectric-mode cavity, this can be achieved by lowering
the band-edges of each modulated mirror segment in a way that linearly increases the
mirror strength. Moreover, the optimal cavity
length for this type of cavity is $L=0$ from Figure.\ref{model}(b),
which simultaneously minimizes the mode volume ($V$) of the cavity
and thus increases $Q/V$. It is worth noting that $L/a=3$ in
Figure.\ref{model}(b) corresponds to the so-called L3
cavity\cite{Noda03}.

One way to
make "modulated Bragg mirror" is to linearly decrease the filling
factor(FF) in nanobeam cavity, which, as we show later, correspond
to a linear increase in mirror strength around the cavity center. FF
is defined as the ratio between the hole area and the area of the unit cell.
Other types of modulation may exist that enable a linearly increasing mirror strength.
In such structure, cavity resonance frequency is expected to be very close
but slightly smaller than the dielectric band-edge of the the
central segment. This is due to the reduction of the hole size in the modulated Bragg mirror, and can be estimated
using perturbation theory. The difference decreases as the number of modulated mirror segments increases.
To summarize the design principles we obtained so far:
(i) zero cavity length ($L=0$), (ii) the length of each segment be
the same (period=a) result in constant phase velocity at $\pi/a$
and (iii) a modulated Bragg mirror results in Gaussian shaped field
attenuation.

\begin{figure}
\centering
\includegraphics[width=9cm,height=8.2cm]{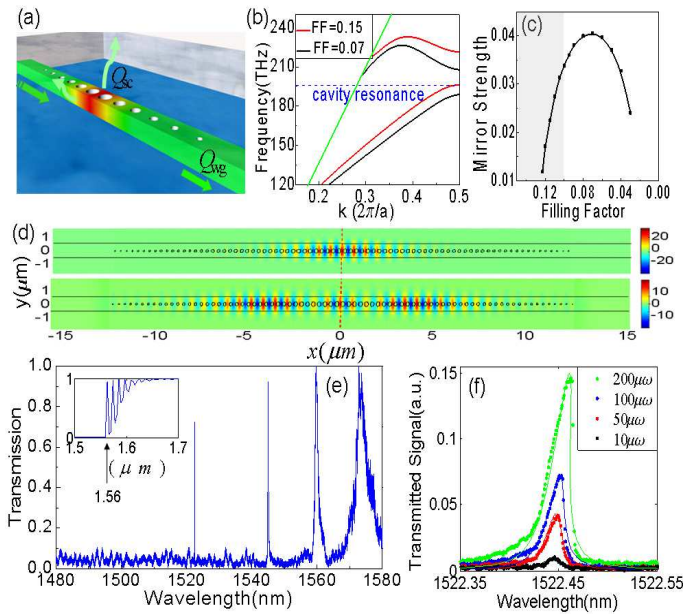}
\caption{\small (a)Schematic of modulated nanobeam cavity.
(b)TE banddiagram of the above cavity with FF=0.15 and FF=0.07. The
resonance of cavity "mod40" is about 1.5\% lower than the
dielectric bandedge of the central section with FF=0.15, due to
modulation. (c)Mirror strength at different FFs. A linear increase
can be seen at cavity center. (d)Simulated Ey profile at the middle
of the naobeam cavity for the fundamental mode(Q=2.7E6) and second mode(Q=8E4). (e)Experiment transmission spectrum of the above mod40. Inset is the
simulation result of the bandedge modes of the current structure. Due to very large photon life time of our ultra-high Q
cavity ($\tau_{\mathrm{photon}}=Q/\omega\sim1\mathrm{ns}$), it becomes nearly
impossible to model transmission through the cavity using 3D FDTD.
Hence the high Q cavity mode does not appear in the simulated
spectrum. (f)Zoom in of the transmitted signal of the fundamental
mode at different input power levels. Dots are experimental data and
lines are fitted curve using nonlinear equation (\ref{transmission}). The power levels in
the legend is the power at the output of the fiber tip. Power coupled to the cavity is smaller due to the
spot-size converter.}\label{3D}
\end{figure}

We demonstrate the power of the recipe by designing an ultra-high Q
and small mode volume PhC nanobeam cavity that operates at $1.525\mu
m$(196.6THz) in a realistic geometry. We assume that the nanobeam is
made with silicon-on-insulator material with 220nm thick Si device
layer (constrained by our SOI wafer properties), and capped with
silica: nanobeam is made of silicon (n=3.46), while holes, bottom
and top cladding are made of material with (n=1.45) (air-holes
backfilled with silica). Our design approach is as follows: (i)We
choose the period a by selecting $n_{\mathrm{eff}}$ to be between
$n_{\mathrm{Si}}=3.46$ and $n_{\mathrm{clad}}=1.45$.
$n_{\mathrm{eff}}=2.5$ is a good compromise, resulting in
$a=\lambda_0/2n_{\mathrm{eff}}\sim300nm$. We note, however, that any
$n_{\mathrm{eff}}$ (and thus any periodicity) which opens a band gap
can be used as a starting point to realize a high Q cavity,
assuming that sufficiently slow modulation is used. (ii) Next, we
choose the width of the nanobeam to be as wide as possible in order to push the mode away from the
light-line, while still being single mode.
Band diagram simulation shows that 700nm is an optimal
choice, that keeps the second order mode (of the same symmetry as the
fundamental one) at the edge of the banggap. (iii) Next, we find the
proper FF which produces a dielectric band edge at $1.525\mu m$. With
$n_{\mathrm{eff}}= 2.5$, FF can be estimated using
$1/n^2_{\texttt{eff}}=(1-\mathrm{FF})/3.46^2+\mathrm{FF}/1.45^2$\cite{sakoda}, which gives 0.19. Then
numerical band diagram simulation shows the actual $\mathrm{FF}=0.15$. Figure.\ref{3D}(b)\&(c) shows the banddiagram and mirror
strength at several FFs. The linear
region between FF=0.15 and FF=0.09 can be used to construct the "modulated Bragg
mirror". (iv) Finally, the nanobeam cavity
needs to be strongly coupled to the feeding waveguide in order to achieve a
large transmission efficiency when probing the cavity. The transmission
at the cavity resonance can be written as
$Q_{\mathrm{wg}}^2/Q^2_{\mathrm{total}}$\cite{johnson}. Therefore,
the cavity $Q$ should be limited by $Q_{\mathrm{wg}}$. The linear decrease of the FF from 0.15 to 0 provides a natural way to
achieve this. More sophisticated couplers will be discussed elsewhere.
In the proposed design method, (i) - (iv), all cavity
parameters are determined using fast band diagram calculations,
only. We note, that the most critical part of our method is that the
modification of the periodic photonic crystal is achieved by keeping
the periodicity $a$ constant, modulating the filing fraction and
using zero cavity length. This approach preserves the phase velocity
of each segment and is essential for realization of high-Q
cavities. Our design strategy has an additional important advantage
over other types of photonic crystal nanobeam
cavities\cite{Noda03}-\cite{Notomi06},\cite{Deotare08}-\cite{painter}:
it provides a natural way to efficiently over-couple the cavity to
the feeding waveguide. Again, we stress that these high Qs are
obtained by-design, performing only simple band-diagram
calculations, and no additional parameter-search trial-based method
was performed. This significantly reduces the computational time to
a few minutes, while a full FDTD simulation takes more than 24 hours
using a grid with 64 processors.

\begin{table}[htbp]
\caption{3D simulation result of waveguide-coupled cavity with 40
and 50 modulated grating sections. The FF is changed from 0.15 at
the center of the cavity to 0. $Q_{\mathrm{wg}}$ and
$Q_{\mathrm{sc}}$ refer to the coupling Q factor to the feeding
waveguide and the scattering Q to the radiation loss.
$V_{\mathrm{eff}}$ is the mode volume normalized by
$(\lambda_0/3.46)^3$.
\\}
\begin{ruledtabular}
\begin{tabular}{cccccccc}
type& $\lambda$($\mu m$) & $Q_\mathrm{sc}$ & $Q_{\mathrm{wg}}$ & $Q_{\mathrm{total}}$ & $V_\mathrm{eff}$ & Trans. \\
\hline mod40 & 1.552 & 2.2E7  & 1.3E6 & 1.2E6 & 1.1 & 0.91\\
 mod50 & 1.549 & 2.5E7  & 2.2E7 & 1.2E7 & 1.2 &0.53\\
\end{tabular} \label{3Dresult}
\end{ruledtabular}
\end{table}

To verify our designs, 3D FDTD modeling is used to study the cavity
with above mentioned parameters. The results are summarized in
Table.\ref{3Dresult} for two cavities with 40 and 50 modulated
mirror segments on each side. It can be seen that cavities feature ultra-high Qs
(both $Q_{\mathrm{sc}}$ and $Q_{\mathrm{wg}}$). At the same time,
high transmission, particularly in cavity "mod40", is obtained. In
principle, Qs can be arbitrarily high by applying sufficient slow
modulation, at the cost of larger mode volumes. However, in
practical case, cavity Q is limited by material losses and
fabrication precisions, and therefore, Q of tens of millions is
sufficient.

To experimentally verify our designs, we fabricated
waveguide-coupled  cavity "mod40" using a silicon-on-insulator(SOI)
wafer with a 220nm device layer on a 2um buried oxide, using the
fabrication procedure described in our previous
work\cite{Deotare08}. Spot size converter\cite{IBM}, that consists
of $2\mu m \times 2\mu m$(cross section) polymer pad($n_{\mathrm{pad}}=1.58$) was used to
couple light in and out of the cavities. The devices were embedded
in polymer with refractive index $n_{\mathrm{clad}}=1.34$. The reduction of
refractive index of top cladding from the simulated one ($n=1.45$)
slightly affects our cavity design. We modeled devices with polymer
top cladding and found that the cavity resonance is shifted to
shorter wavelength ($\lambda_{\mathrm{model}}=1517nm$), the cavity $Q_{\mathrm{total}}$
is 2.72E6, and the the on-resonance transmission is $T=75\%$.

We characterized the device using a tunable laser
source(1470-1580nm) and a tapered fiber tip(2.5$\mu$m spot diameter)
to couple light in and out of the polymer waveguide. A polarizer is
placed at the output to filter out the TM-like mode.
Figure.\ref{3D}(e) is the experiment transmission spectrum through one of
the resonators. The signal is normalized by the band-edge modes,
which has a unity transmission as verified by 3D FDTD simulations
(inset).  A nonlinear bistability is
observed and Figure.\ref{3D}(f) further shows the zoom in of the
fundamental mode at different input power levels. We fitted the
experimental data using the following expression typical for the
nonlinear bistability
\begin{equation}
T=\frac{P_{out}}{P_{in}}=\frac{Q^2_{\mathrm{total}}/Q^2_{\mathrm{wg}}}{1+(P_{out}/P_0-2(\lambda-\lambda_0)/\gamma_0)^2}
\label{transmission}\end{equation}
$P_0=3\kappa Q_{\mathrm{total}}Q_{\mathrm{wg}}(\omega/(2nc))^2\chi^{(3)}$ is the characterized power
in the presence of a third order nonlinearity.  $\kappa$ is the
nonlinear feedback parameter introduced by Soljacic et.
al\cite{soljacic02}. $\kappa \sim 1/V_{\mathrm{eff}}$ is an
indicator of the extent of the field that is confined in the
nonlinear region. $\gamma_0$ is the natural cavity linewidth. We
obtained a $Q=80,000$ and an on-resonance transmission $T=73\%$ for
the on-substrated and capped cavity. This corresponds to a
$Q_{\mathrm{sc}}=500,000$, which is comparable with our previously
reported results for free-standing photonic crystal nanobeam
cavity\cite{Deotare08}. The demonstrated transmission is much higher
than previous works with the same Q\cite{Notomi09}.

In summary, we proposed and demonstrated a deterministic design of
the high Q PhC nanobeam cavities. Our cavities are ideally suited
for the coupling to the feeding waveguides and allow for high
transmission efficiency. This makes them ideal candidates for the
realization of densely integrated photonic systems, and are suitable
for applications ranging from optical interconnects to biochemical
sensors. Finally, record high Q/V ratios of photonic crystal
nanobeam cavities will enable further fundamental studies in
spontaneous emission control, nonlinear optics and quantum optics.

\begin{acknowledgments}
We acknowledge useful discussions with M.W. MuCutcheon. This work
is supported by NSF Grant No. ECCS-0701417 and NSF CAREER grant. Device fabrication is
performed at the Center for Nanoscale Systems(CNS) at Harvard.
\end{acknowledgments}

\bibliographystyle{unsrt}

\end{document}